# Physics Archives
*November 2009*

*Mapping Complex Networks:*
*Exploring Boolean Modeling of Signal Transduction Pathways*

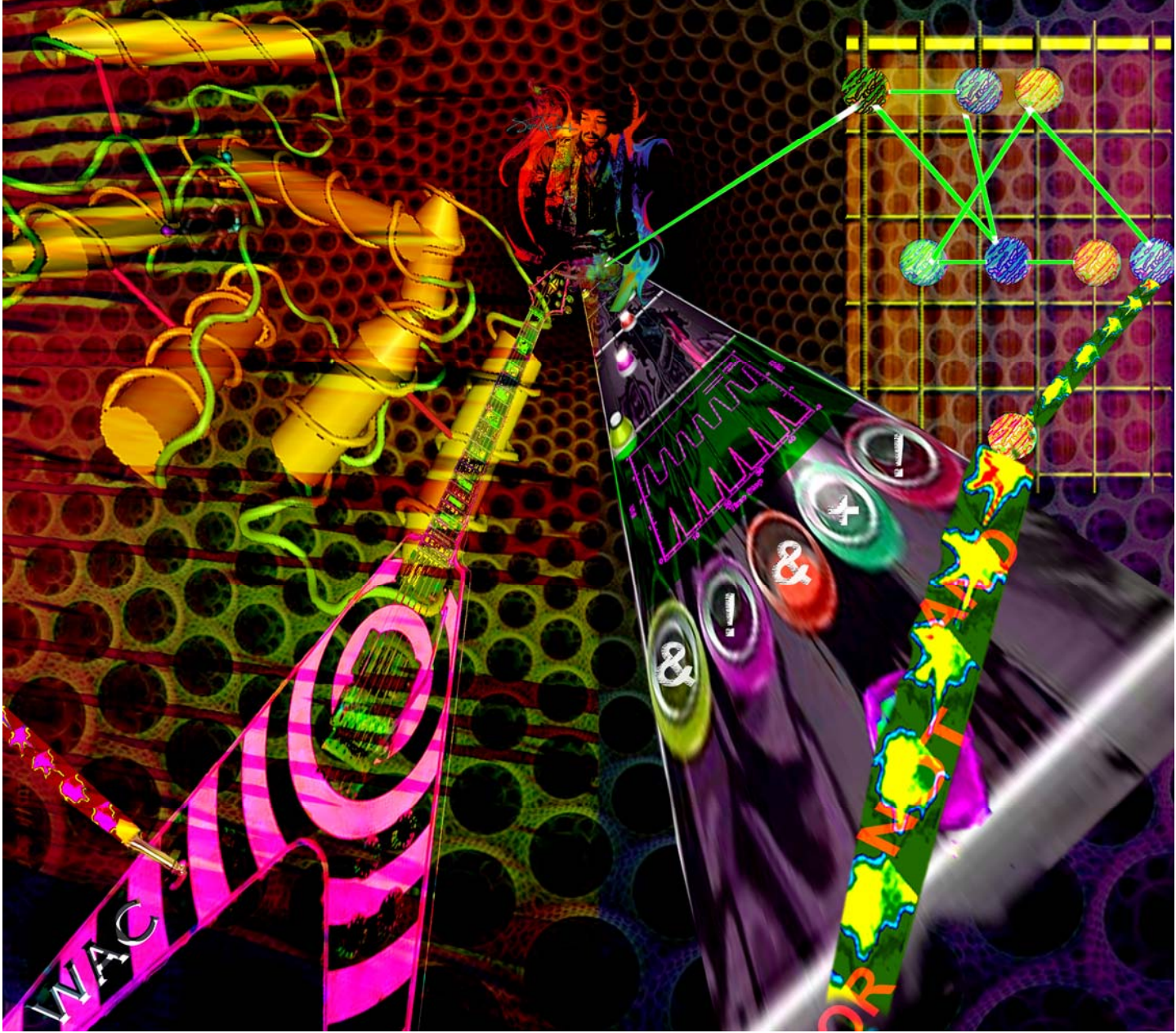

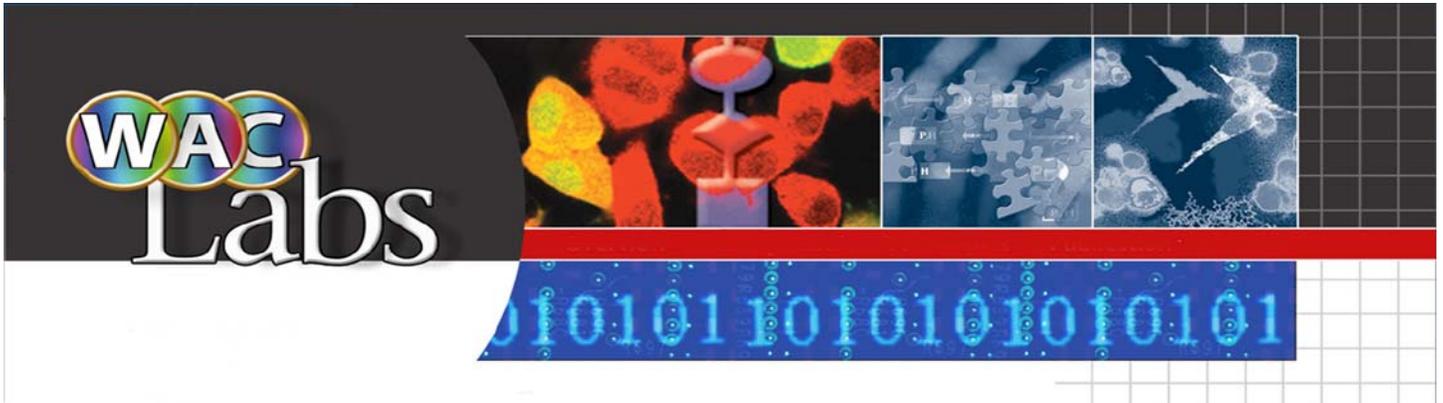

# Mapping Complex Networks: Exploring Boolean Modeling of Signal Transduction Pathways

Gaurav Bhardwaj[1,2], Christine P. Wells[1], Reka Albert[2,3], Damian B. van Rossum[1,2,*], and Randen L. Patterson[1,2,*]

(1) Department of Biology, The Pennsylvania State University, 16801
(2) Center for Computational Proteomics, The Pennsylvania State University, 16801
(3) Department of Physics, The Pennsylvania State University, 16801

[*]Address correspondence to:

Randen L. Patterson, 230 Life Science Bldg, University Park, PA 16802. Tel: 001-814-865-1668; Fax: 001-814-863-1357; E-mail: rlp25@psu.edu.

Damian B. van Rossum, 518 Wartik Laboratory, University Park, PA 16802. Tel: 001-814-865-1668; Fax: 001-814-863-1357; E-mail: dbv10@psu.edu.

## Abstract

In this study, we explored the utility of a descriptive and predictive bionetwork model for phospholipase C-coupled calcium signaling pathways, built with non-kinetic experimental information. Boolean models generated from these data yield oscillatory activity patterns for both the endoplasmic reticulum resident inositol-1,4,5-trisphosphate receptor ($IP_3R$) and the plasma-membrane resident canonical transient receptor potential channel 3 (TRPC3). These results are specific as randomization of the Boolean operators ablates oscillatory pattern formation. Furthermore, knock-out simulations of the $IP_3R$, TRPC3, and multiple other proteins recapitulate experimentally derived results. The potential of this approach can be observed by its ability to predict previously undescribed cellular phenotypes using *in vitro* experimental data. Indeed our cellular analysis of the developmental and calcium-regulatory protein, DANGER1a, confirms the counter-intuitive predictions from our Boolean models in two highly relevant cellular models. Based on these results, we theorize that with sufficient legacy knowledge and/or computational biology predictions, Boolean networks provide a robust method for predictive-modeling of any biological system.

**Introduction**

Calcium signaling networks are comprised of multiple nodes including: (i) receptors and ligands, (ii) soluble second messengers, and (iii) selective/non-selective ion-channels. Within these signal transduction pathways, ligand-binding to either a G-protein coupled receptor (GPCR) or tyrosine-kinase receptor (RTK) in the plasma-membrane initiates numerous short-term and long-term cellular signals. These signals initiate cellular programs responsible for growth, development, secretion, and apoptosis [1;2]. Subsequent to ligand binding, GPCRs activate the heterotrimeric G-protein complex (G$\alpha$ and G$\beta\gamma$) which stimulates phospholipase C-beta (PLC$\beta$) mediated catalysis of phosphatidylinositol (4,5) bisphosphate (PIP$_2$) to the second messengers inositol-1,4,5-trisphosphate (IP$_3$) and diacylglycerol (DAG). Similarly, ligand activation of RTKs results in the direct activation of phospholipase C-gamma (PLC$\gamma$), and its catalysis of PIP$_2$. In this way, IP$_3$ is generated and activates inositol-1,4,5-trisphosphate receptors (IP$_3$R), a large conductance Ca$^{2+}$ channel on membranes of Ca$^{2+}$ containing stores. Active IP$_3$R releases Ca$^{2+}$ into the cytosol, as well as other cellular compartments (e.g. nucleus, mitochondria, etc)[1;3]. Recent studies also demonstrate that IP$_3$Rs are regulated by a large suite of proteins which can both positively and negatively influence their calcium conductance. Further, IP$_3$R is also integral to Ca$^{2+}$ entry mechanisms [1;4], and is a known regulator of transient receptor potential channels (TRPs), a superfamily of ion-channels involved in sensory perception [4;5].

The output of these signals provides Ca$^{2+}$ oscillations/waves which spatially regulate protein function to accomplish diverse physiological/pathophysiological processes, such as sensory perception [6], cell development and growth[7;8], hearing[9;10], taste[11], and fertility[12] as well as multiple diseases including mucolipodosis [13;14] and polycystic kidney disease[4;13;15]. Although our understanding of many of the core components is well-developed for certain steps within IP$_3$R containing Ca$^{2+}$ networks, our understanding of how these signals are integrated with temporal and spatial precision is severely lacking.

The Ca$^{2+}$ signaling network, even within a single cell, may contain hundreds if not thousands of different nodes (proteins, small molecules, lipids, ions, etc). Pioneers of network analyses have attempted to model Ca$^{2+}$ signaling networks using kinetic data from biochemical experimentation [16-20]. One such study by Bhalla and Iyengar [21] demonstrated that by building a library of small signaling network modules (e.g. PLC, MAP kinase, PKC, PLA$_2$, etc), they could build a network which could simulate multiple functional outputs such as intracellular Ca$^{2+}$ concentration, kinase activity, phosphatase activity, etc. Further, a number of non-intuitive results were derived that could be recapitulated in the laboratory [21]. Kinetic models built from experimental data can also successfully predict the effects of protein knock-down on cellular Ca$^{2+}$ dynamics [18]. Maurya and Subramaniam demonstrated that protein knock-down does not have a linear correlation to loss of function (e.g. 18% knock-down of G$\beta\gamma$ equals a 50% loss of activity, which is also equivalent to a 60% knock-down of receptor protein)[17;18]. Taken together, these results highlight the utility of kinetic network modeling.

Network analyses using quantitative kinetic data are powerful and predictive[17;18;22;23]; however, the differential equations required to process these types of data require detailed information and enormous computational power. Furthermore, kinetic data for individual steps in the network is often lacking. Thus, generating larger scale network simulations is sometimes not feasible. Discrete network modeling (e.g. Boolean) may provide a solution to this problem, as it was demonstrated that in many networks kinetic parameters are not essential in describing the overall dynamics [24;25]. Moreover, Boolean models characterize nodes with two qualitative states and only use three operators (AND, OR, NOT); thus, significantly reducing the computational power needed to run Boolean simulations as compared to kinetic network models [26;27].

In the present study, we investigate whether Boolean modeling can be used to study PLC-coupled $Ca^{2+}$ signaling pathways. Our results demonstrate that (i) $Ca^{2+}$ networks can be constructed from legacy knowledge, (ii) experimental data from these multiple sources can be used to define network directionality and Boolean update rules for each node in network, and (iii) network modeling based on these Boolean rules provide a descriptive and predictive model of PLC-mediated $Ca^{2+}$ signaling. In addition, we experimentally validate a previously uncharacterized node (DANGER1a) in the network simulations. Taken together, we suggest that Boolean modeling provide an effective method for creating receptor specific and/or cell-specific signaling network models.

## Results
### Constructing the Bionetwork and Generating Boolean Rules

We chose to build a model for PLC-coupled $Ca^{2+}$ signaling pathways due to the significant literature base for this receptor [28]. Nodes within the network were chosen based on two criteria: (1) they have known activity(s) within these pathways and (2) they are known to have causal interactions, and not just associative interactions with other nodes in the network (see Methods for a complete description). Figure 1 depicts the bionetwork generated using this scheme. Although this is a subset of the known nodes/integrators for phospholipase C-mediated signaling, bionetworks are often resilient, even in the absence of a significant number of nodes [29]. In addition, nodes were chosen that have, in general, considerable experimental support[1;3]. Furthermore, as all GPCRs and RTKs flow through either PLCβ or PLCγ, our network results

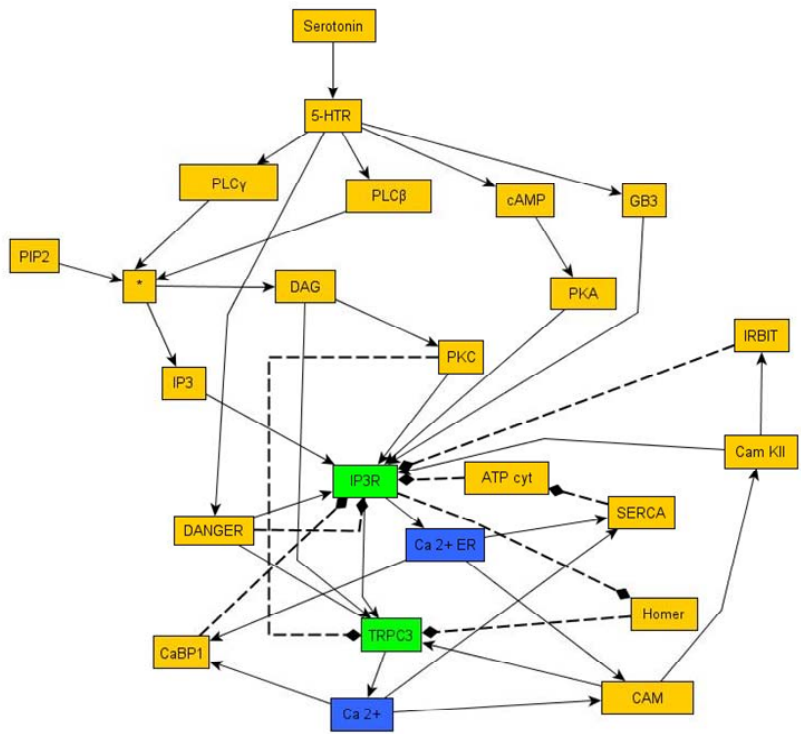

**Figure 1**: **Bionetwork of Serotonin Induced Calcium Signaling Pathway**
Calcium channels have been shown in green color, calcium release and entry are blue, and all other signaling components have been shown in yellow. Solid arrows represent positive interaction between nodes, and dashed arrows represent inhibition. '*' denotes hydrolysis of phosphatidylinositol bisphosphate. Full names of abbreviations used in the network are: 5-HTR, 5-hydroxytryptamine receptor; PLCβ/γ , Phospholipase C Beta/Gamma; PIP2, phosphatidylinositol bisphosphate; DAG, Diacylglycerol; IP3, Inositol trisphosphate; PKA, Protein Kinase A; PKC, Protein kinase C; SERCA, Sarco/Endoplasmic Reticulum Ca2+-ATPase; IP3R, Inositol 1,4,5-Trisphosphate Receptor; TRPC3, Transient receptor potential cation channel, subfamily C, member 3; CaBP1, Calcium Binding Protein1;ATPcyt , cytoplasmic ATP in vicinity to IP3R channels; CAM, Calmodulin; CamkII, Ca2+/calmodulin-dependent protein kinase II should be applicable to numerous receptor types (e.g. muscarinic, serotonergic, B-cell receptors, etc). From this topology, we observe that the IP$_3$R and the canonical TRP channel 3 (TRPC3) form hubs (highly-connected nodes) within the network. This implies that these channels are highly regulated, which is reasonable as both of these proteins flux the signaling ion, $Ca^{2+}$.

After constructing the network, we defined our Boolean rules for updating each node in terms of logic operators (AND, OR, NOT) (Figure 2). In this method, nodes can have only two states, ON or OFF (see Table 1). ON can represent states such as ACTIVE (e.g. kinases, phosphatases, lipases etc.), OPEN (e.g. channels) and/or BINDING (e.g protein-protein or protein-lipid interactions). Conversely, OFF can represent INACTIVE, CLOSED and/or NON-BINDING states.

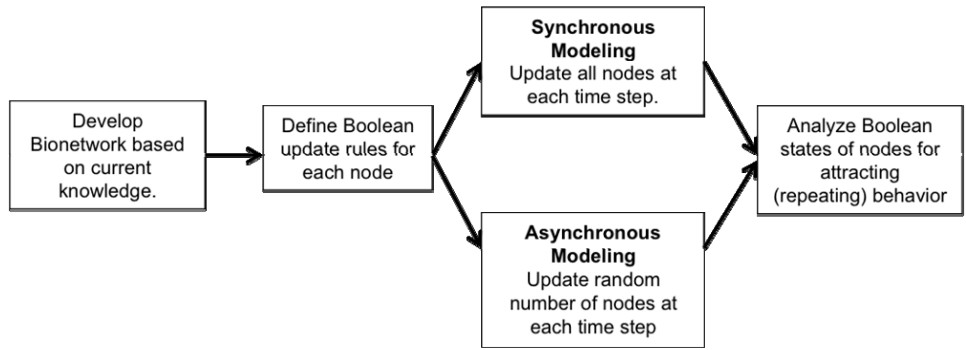

**Figure 2: Flowchart of Bionetwork-Boolean Modeling Workflow**
Bionetwork is developed based on current knowledge and rules for updating each node defined in form of logical operators (AND/OR/NOT). For synchronous modeling, each node is updated at each time step. For asynchronous modeling, random numbers of nodes are updated at each time step. Boolean states of each node in each time step is noted down in each update method and analyzed for attracting behavior. (see Methods for details)

To define the Boolean update rules for each node, directionality information between nodes, as provided by legacy knowledge, is converted to logical operator driven statements. Thus, the future state of each node is dependent upon the present state of its interacting nodes. We created these logic operators for all nodes in the network using either legacy knowledge or our own experimental data in defining rules (see Table 1 for rules used). For example, we used our own experimental results to define update rules for

**Table 1: Boolean Rules for updating each node in network**

| Node | Boolean Regulatory Rule |
|---|---|
| HTR | $HTR^*$ = serotonin |
| PLCβ | $PLCβ^*$ = HTR |
| cAMP | $cAMP^*$ = HTR |
| DAG | $DAG^*$ = PIP$_2$ and (PLCβ or PLCγ) |
| PLCγ | $PLCγ^*$ = HTR |
| IP$_3$ | $IP_3^*$ = PIP$_2$ and (PLCβ or PLCγ) |
| PKC | $PKC^*$ = {DAG and ($Ca^{2+}$ or $Ca^{2+}_{ER(Low)}$ or $Ca^{2+}_{ER(High)}$)} |
| TRPC3 | $TRPC3^*$ = {(not HOMER) and (IP$_3$R$_{Low}$ and PLCγ and CAM)} or (DAG and not PKC) |
| HOMER | $HOMER^*$ = not IP$_3$R$_{Low}$ |
| $Ca^{2+}_{ER(High)}$ | $Ca^{2+*}_{ER(High)}$ = IP$_3$R$_{High}$ |
| CAM | $CAM^*$ = ($Ca^{2+}$ or $Ca^{2+}_{ER(Low)}$ or $Ca^{2+}_{ER(High)}$) |
| SERCA | $SERCA^*$ = ($Ca^{2+}$ or $Ca^{2+}_{ER(Low)}$ or $Ca^{2+}_{ER(High)}$) |
| CaBP1 | $CaBP1^*$ = $Ca^{2+}_{ER(High)}$ and DANGER |
| ATP$_{cyt}$ | $ATP^*_{cyt}$ = not SERCA |
| CamKII | $CamKII^*$ = CAM |
| IP$_3$R$_{Low}$ | $IP_3R^*_{Low}$ = {(IP$_3$ and DANGER and (PKC or PKA or CamKII or Gβ3)} and not (IRBIT or CaBP1 or ATP$_{cyt}$) |
| IP$_3$R$_{High}$ | $IP_3R^*_{High}$ = {(IP$_3$ and DANGER and PKC and PKA and Gβ3)} and not (IRBIT or CaBP1 or ATP$_{cyt}$) |
| DANGER | $DANGER^*$ = HTR |
| Gβ3 | $Gβ3^*$ = HTR |
| IRBIT | $IRBIT^*$ = CamkII |
| $Ca^{2+}$ | $Ca^{2+*}$ = TRPC3 |

Heterotrimeric G-protein β3 (Gβ3). Studies from Zeng et al. [30] demonstrate that heterotrimeric G-proteins have the ability to stimulate IP$_3$R activity directly, although direct binding has not been determined. We tested the hypothesis that Gβ3 and IP$_3$R bind directly. Yeast two-hybrid analysis reveals that the third and fourth WD40 repeats of Gβ3 are critical for binding (Figure 3a). We next examined the direct binding of the IP$_3$R to full-length Gβ3 or a mutant Gβ3 containing deletions in the third and fourth WD40 repeats (aa166–207) (Figure 3b). Whereas full-length Gβ3 binds to IP$_3$R, binding is abolished by deletion of the two WD40 repeats.

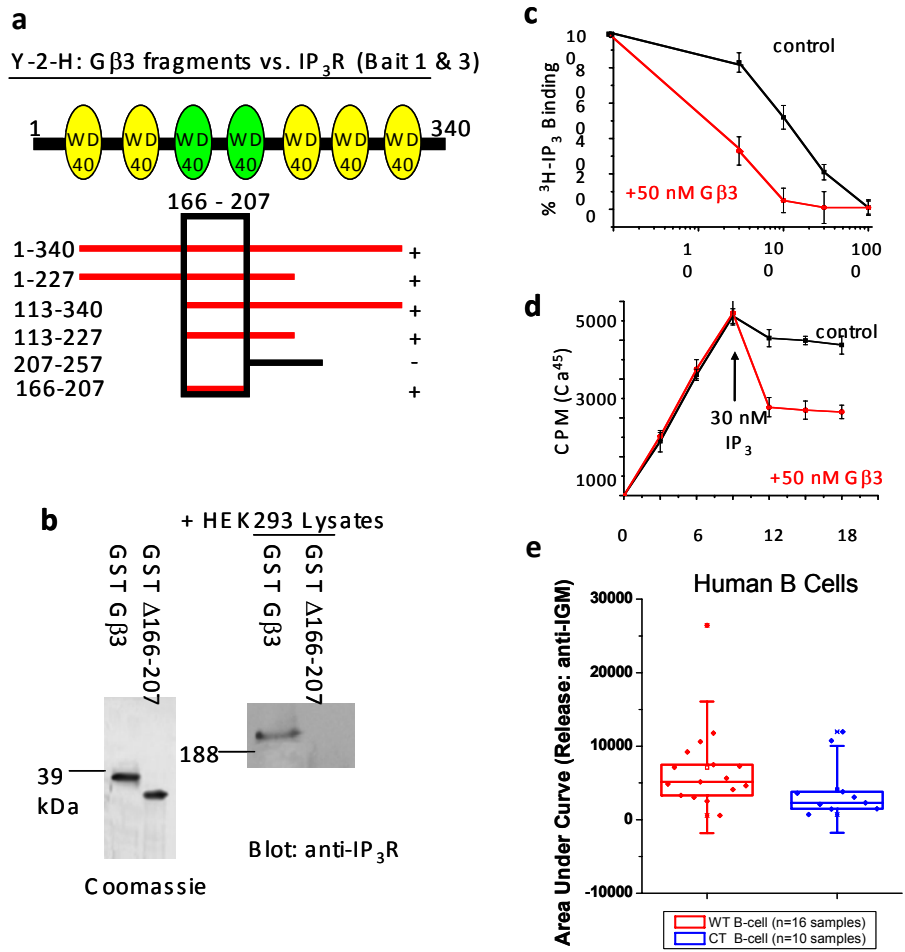

To explore an influence of Gβ3 on IP$_3$R function, we first evaluated the influences of 50 nM Gβ3 on [$^3$H]IP$_3$ binding to IP$_3$R in rat brain membranes (Figure 3c). Gβ3 increases the potency of unlabeled IP$_3$ in competing for [$^3$H]IP$_3$ binding ≈3 fold. We next evaluated the influence of 50 nM Gβ3 on IP$_3$-induced Ca$^{2+}$ release from microsomal membrane preparations of COS7 cells (Figure 3d). We do not observe activation of IP$_3$R by Gβ3 alone, even with high concentration of Gβ3 (10 μM) (data not shown). These observations suggest that Gβ3 acts by enhancing the affinity of IP$_3$ for IP$_3$R rather than by altering the channel itself, which would lead to a greater maximal release. Taken together, the activity results are in accord with Zeng et al. although we do not observe direct activation of the IP$_3$R by Gβ3 in the absence of IP$_3$.

Further, we generated primary B-cell lines from human patients which were either WT or homozygous for the C825T polymorphism [31]. This is a naturally occurring polymorphism which results in a frameshift mutation leading to the deletion of amino acids 166-207 in Gβ3. In these cells we measured Ca$^{2+}$ release in response to B-cell receptor stimulation with anti-IgM which stimulates the RTK B-cell receptor[32-34]. The area under the curve from these measurements were quantified from numerous replicates and compared (see Methods for complete description). Although we do observe that some of these preparations have alterations in $Ca^{2+}$ signaling, our results from multiple genetic backgrounds provided no statistical difference between WT and mutant preparations (Figure 3e).

**Figure 3**: **Gβ3 binds and regulates IP3R activity** (a) Yeast two-hybrid analysis shows that Gβ3 binds to IP3R through its 3rd and 4th WD repeats (aa 166-207). Fragments of Gβ3 lacking in region 166-207 do not show any binding (b) Pull down experiments in HEK293 lysates with GST tagged full length and mutated Gβ3 (Δ166-207) were performed. Full length Gβ3 shows a direct binding to IP3R while there is no binding between mutant Gβ3 and IP3R. (c) In [3H]IP3 treated Rat brain membranes, Gβ3 (50nM) increased IP3 binding to IP3R. (d) In microsomal membrane of COS7 cells, Gβ3 (50 nM) leads to maximum calcium release at 30 nM IP3. (e) Quantification of Ca2+ mobilization in human B cells lines from either Wild Type (WT) or C825T homozygous mutants stimulated with 10 mM anti-human IgM. Area under individual calcium traces was measured and it shows that there is no statistical difference in Ca2+ release between WT and C825T mutants (Area under the curve, n= average 396 cells from 16 replicates across 4 different cell lines for WT and n=average 220 cells from 10 replicates across 3 different cell lines for C825T mutants, Error Bars: Std. Dev.)

Based on these experimental findings we defined the Boolean rule for the Gβ3 node (Figure 1 and Table 1). Specifically, in our rules Gβ3 enhances the activity of the $IP_3R$ although it is non-essential. Indeed, the $IP_3R$ is expressed in all cells[1;35-37] whereas Gβ3 is not [38].

## Synchronous and Asynchronous Simulations of the Serotonin Bionetwork

Following, we performed two types of Boolean modeling. Synchronous Boolean modeling assumes that the time required for each biological reaction is comparable and thus a unit time-step can be defined (Figure 2). Therefore, when performing synchronous modeling we simultaneously update all node states according to the rules (Table 1) at each time-step. Conversely, asynchronous Boolean modeling assumes that each biological reaction has distinct durations. Since there is not enough quantitative information available to estimate the reaction durations, we randomly shuffle the order of nodes at each time-step (Figure 2). Thus, this approach samples differential time intervals for each biological reaction in network. The results from these simulations were assayed for their attracting behavior (i.e. repetitive behavior).

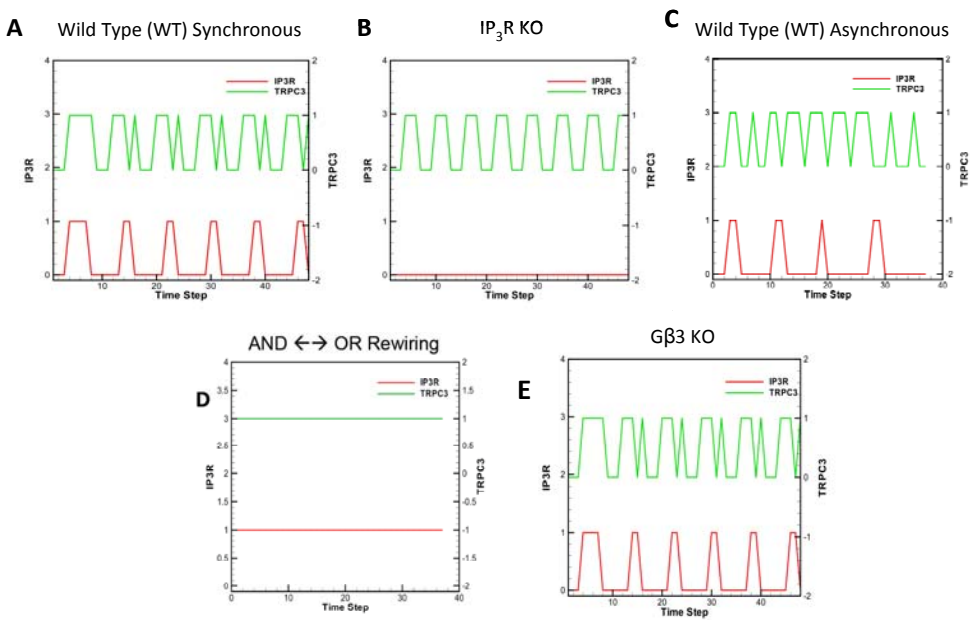

Results from our simulations can be plotted as a function of the Boolean state versus the time-step. Figure 4A depicts the behavior for the $IP_3R$ and TRPC3 nodes in the network. We observe that both proteins have oscillatory behavior, in agreement with experimental data [39]. Interestingly, these proteins display different oscillatory patterns in terms of their time-steps. As a control, we knocked-out the $IP_3R$ from the bionetwork and observe that the $IP_3R$ has no activity; interestingly, TRPC3 is still oscillatory, but has an altered pattern of oscillations (Figure 4B). This may be expected as TRPC3 can be activated in an IP$_3$R-independent manner, although this has often been described in over-expression experiments [5]. In our asynchronous model (Figure 4C), we still observe an oscillatory pattern, although both IP$_3$R and TRPC3 oscillatory patterns no longer exhibit a fixed time period. This demonstrates that this network is robust in generating oscillatory patterns. Moreover, since we are trying to find a correlation between change in oscillatory time periods and change in cellular Ca$^{2+}$ levels, asynchronous Boolean modeling poses a major limitation. Specifically, asynchronous modeling alters the rate at which nodes are updated; thus, the time-periods measurements between oscillations would be corrupted. In order to characterize the oscillations of the system we need a non-random model; therefore, for the purposes of this manuscript we chose to explore synchronous models. Importantly, in control experiments, we randomized the Boolean rules for all nodes by changing all ANDs to OR and vice versa, and observe no oscillatory patterns in our synchronous models (Figure 4D). This data confirms the oscillatory patterns observed under both synchronous and asynchronous simulations are not random in nature.

**Figure 4: Boolean Modeling Predicts Wild-Type (WT) and Knock Out (KO) phenotypes**
**(A)** Plots of Boolean state vs. Time steps from synchronous modeling recapitulate the oscillatory pattern of IP3R and TRPC3 in Wild Type cells. Both the channels show a distinct oscillatory pattern with fixed time periods. **(B)** Boolean modeling with IP3R node knocked out (Control) shows that IP3R Boolean state remains fixed at 0 (Off) and TRPC3 oscillatory pattern changes predicting a reduced Ca2+ entry. **(C)** Asynchronous update method replicates the oscillatory pattern of IP3R and TRPC3 but random update of nodes leads to varying time periods. **(D)** Rewiring of AND & OR in logical rules fails to provide oscillations in the model. **(E)** Boolean modeling of Ca2+ network after knocking out Gβ3 node show no change in oscillatory pattern of IP3R and TRPC3 Boolean states as compared to Wild Type (WT)

**Table 2: Knock out Phenotypes Predicted by Boolean Modeling and Literature Predictions**

| Knock Out Node(s) | Oscillation Pattern | | Predicted Phenotype | References |
|---|---|---|---|---|
| | IP$_3$R | TRPC3 | | |
| DANGER | 1(ON)-6(OFF) | 3(ON)-4(OFF) | IP$_3$R remain open for shorter interval and TRPC3 oscillation change to single type oscillations | Experimentally validated |
| IP$_3$R | 0 | 3(ON)-3(OFF) | Calcium release stops and TRP channels show only DAG mediated/Basal oscillations. | [1;3;5;63] |
| GB3 | 2(ON)-6(OFF) | 3(ON)-3(OFF)-1(ON)-1(OFF) | No change as compared to wild type | Experimentally validated |
| CamKII | 3(ON)-5(OFF) | 5(ON)-3(OFF) | Calcium release and entry both increase as channel remain open for longer time and closing time decreases. | |
| PKC | 0 | 1 | Calcium release goes off completely and TRP channels remain constantly 'ON' | [5;64] |
| ATP$_{cyt}$ | 3(ON)-6(OFF) | 3(OFF)-3(ON)-1(OFF)-2(ON) | Calcium release increases, as IP$_3$R remain open for longer times. TRPC also remain open longer as compared to wild type. | [65;66] |
| CaBP1 | 2(ON)-6(OFF) | 3(ON)-3(OFF)-1(ON)-1(OFF) | No change as compared to wild type | [67] |
| IP3 | 0 | 3(ON)-3(OFF) | Same as IP$_3$R KO | [68] |
| SERCA | 3(ON)-6(OFF) | 3(OFF)-3(ON)-1(OFF)-2(ON) | Similar to ATP-vicinity KO | |
| DAG | 4(ON)-6(OFF) | 2(ON)-8(OFF) | Calcium release affected, as channels remain open for longer time but there is no high concentration release. Entry decreases, as TRPC remains closed for longer times. | [5;64] |
| PKA, cAMP | 2(ON)-6(OFF) | 3(ON)-3(OFF)-1(ON)-1(OFF) | Calcium release decreases as channel remain open for same time as wild type but no high concentration release events. Entry remains normal. | [69-71] |
| Ligand, Receptor, PLC G, PLC B | 0 | 0 | No oscillations in network as it remains all 'OFF' | [28;72-77] |
| IRBIT | 3(ON)-5(OFF) | 5(ON)-3(OFF) | Release and entry both increase | [42;43] |
| Calmodulin | 3(ON)-4(OFF) | 3(ON)-4(OFF) | Release increases but entry decreases | [78-80] |
| HOMER | 2(ON)-6(OFF) | 5(ON)-3(OFF) | Release remains unaffected while calcium entry increases | [41] |

**Experimental Validation of Network Simulations**

To determine whether our synchronous models are informative, we knocked-out all of the nodes in the network one at a time and recorded the results (Table 2). Several of these simulations are corroborated by experimental literature. For example, simulated deletion of Homer results in increased Ca$^{2+}$ entry through TRPC3 without altering Ca$^{2+}$ release through IP$_3$Rs, as is observed experimentally [40;41]. Another example is IRBIT, which is an inhibitor of the IP$_3$R. Its absence in the network predicts increased Ca$^{2+}$ release and entry, in accord with the studies of Ando et al [42;43]. Knock-out of PKC from our network predicts that Ca$^{2+}$ release is abolished, while Ca$^{2+}$ entry is substantially enhanced. Indeed, the work of Venkatachalam et al. clearly demonstrates that pharmacological inhibition of PKC validates our prediction [5].

Network simulations with Gβ3 knocked-out from the network, exhibit no change in calcium mobilization (Figure 4E). Based on *in vitro* data, one might predict that removal of Gβ3 from the network would result in a decrease in Ca$^{2+}$ release. Although, our Boolean update rule for Gβ3 was based on *in vitro* data, but when taken together with essentiality information, which was derived from differential and compensatory cellular expression, the rule recapitulates the *in vivo* phenotype (Figure 3e).

## Experimental Validation of Network Predictions
### DANGER1a

To extend on these findings, we examined the role of DANGER1a; a member of the newly-discovered DANGER developmental superfamily [44]. This protein was discovered through a yeast-2-hybrid screen with the IP$_3$R, and was demonstrated to regulate IP$_3$R activity and neuronal development [44;45]. Specifically, DANGER1a decreases the open-probability of the IP$_3$R at high Ca$^{2+}$-concentrations (>300 nM). This *in vitro* data was used to construct the Boolean rules for DANGER1a. Specifically, we defined two different activation states for IP$_3$R release, HIGH and LOW. Under HIGH release conditions, DANGER1a acts as an inhibitor to IP$_3$R activity, while, under LOW release conditions, it plays a non-essential, non-inhibitory role. As DANGER1a is inhibitor in our *in vitro* analyses, one would expect that knock-out of DANGER1a would result in increased Ca$^{2+}$ release. However, as shown in Figure 5a, our dynamic model predicts that knock-out of DANGER1a would result in an overall decrease of Ca$^{2+}$ mobilization.

To reconcile these results we turned to DANGER1a knock-out mice. Shown in Figure 5b, the Cre/Flox DANGER1a knock-out mouse was properly constructed as shown by southern analysis and PCR (Figure 5b, see Methods for complete description). From embryonic mice, we cultured primary spinal cord neurons and measured intracellular Ca$^{2+}$ transients in response to serotonin, using the fluorescent Ca$^{2+}$ indicator, Fura-2AM (see Methods). These experiments reveal that spinal cord neurons from mice, either heterozygous or homozygous for DANGER1a knock-out, have a decrease in overall Ca$^{2+}$ mobilization when stimulated with serotonin (Figure 5c). Surprisingly, these results recapitulate the predictions from our Boolean model, not the intuitive expectations that could be projected from the *in vitro* data alone.

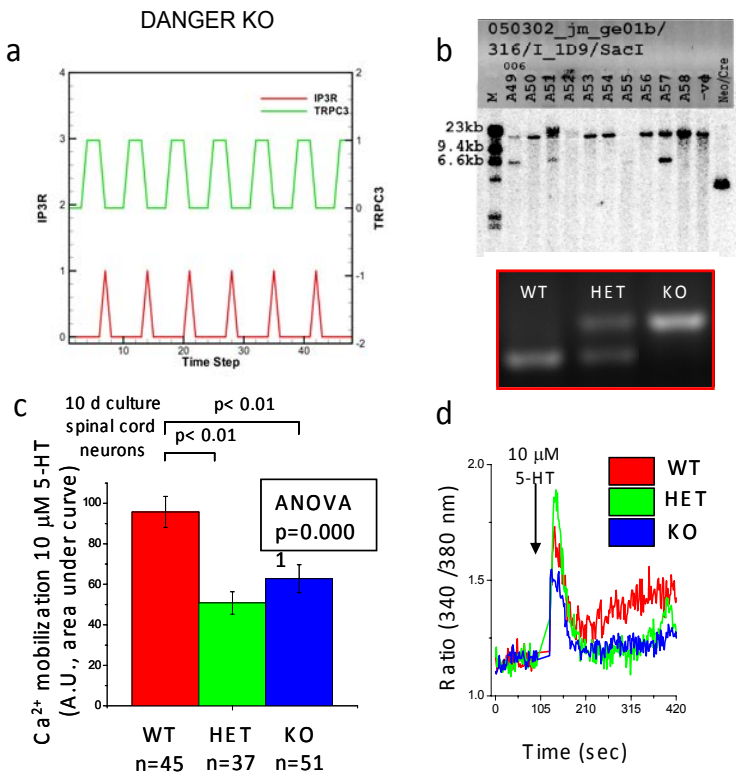

**Figure 5: Boolean Modeling accurately predicts *in vivo* DANGER KO phenotype. (a)** Boolean Modeling of our network with DANGER1a node knocked out predicts that an altered oscillatory pattern and a reduced Ca2+ mobilization. **(b)** (top) Southern analysis of embryonic stem cells generated by Ozgene©. Cells from lane A57 were used to make DANGER1A mice. (bottom) PCR analysis for DANGER1a transcript in Wild Type (WT), Heterozygous (HET) and DANGER1a knock out (KO) mice. **(c)** Quantification of Ca2+ mobilization in spinal cord neurons from WT, HET and DANGER1a KO mice stimulated with 1 mM serotonin (% of WT, area under curve, n = total number of cells from 3 independent experiments, Error bars: Std dev) **(d)** Representative calcium traces from one complete imaging experiment.

**Discussion**

In this study, we successfully used qualitative data from various experimental studies to develop a Boolean dynamic model for the PLC-coupled $Ca^{2+}$ signaling pathway. A number of implications can be derived from this work: (i) descriptive and predictive models can be built in the absence of kinetic data for multiple receptors (muscarinic, serotonergic, B-cell receptor), (ii) these models can be used to predict mutant/knock-out phenotypes which can be validated experimentally, and (iii) in some cases, these models can be used to anticipate non-intuitive biological effects. From these preliminary studies, we propose that refining Boolean models of $Ca^{2+}$ signaling will likely provide a rich resource for experimental biologists.

By gathering and organizing current knowledge about PLC-coupled $Ca^{2+}$ signaling into a bionetwork, the biological information is organized into local structures. This allows the interactions between components of the pathway to become more accessible and easier to understand. Indeed, network topology can identify the various feedback and feed-forward loops, as well as the most connected nodes (i.e. hubs) in the system. Feedback and feed-forward loops represent the regulatory control in the signaling pathway, while hubs signify the most regulated components in the network. In our bionetwork, both the $IP_3R$ and TRPC3 emerge as hubs. This is not surprising as both $IP_3R$ and TRPC3 are heavily regulated calcium channels [1;4;46]. We also observe that our simulations accord with many observations from the literature (Table 2). Importantly, our bionetwork also approximates the oscillatory behavior observed during cellular $Ca^{2+}$ signaling, as well as the experimental results obtained for $G\beta 3$ and DANGER1a in serotonin-induced signaling pathways (Fig 4 and 5). Although similar types of results have been obtained from differential equation based models of calcium signaling [17;18], we propose that Boolean modeling represents an independent and powerful tool for examining signaling pathways when detailed kinetic data does not exist. Due to the need for extensive computational power, kinetic models are often not feasible; thus Boolean bionetwork modeling has great potential for measuring $Ca^{2+}$ signaling pathways that lack kinetic data.

Of particular interest is the observation that our Boolean model creates oscillatory patterns for both $IP_3R$ and TRPC3 channel opening in wild type and various mutant/knock-out conditions (Fig 4a-c,e). This oscillatory pattern is not random and/or an artifact of our approach; when the logical rules for 'AND'/'OR' are rewired, the network fails to generate any oscillations (Fig 4d). The accuracy of our network is further evident from comparisons with experimental results. Changes in oscillatory frequency of calcium channel activity observed in our Boolean analysis correlate with experimental measurements for changes in $Ca^{2+}$ mobilization in various mutant/knock-out cell types (Fig 3-5 and Table 2). The ability of this bionetwork to consistently replicate WT and mutant phenotypes suggests that the data used to construct the rules is accurate and that general trends in $Ca^{2+}$ mobilization can be modeled.

Another interesting finding is that in some cases when *in vitro* and *in vivo* results conflict, Boolean modeling can predict results for the *in vivo* phenotype using *in vitro* rules. In our study, network modeling predicted phenotypes that were counter-intuitive to *in vitro* results for DANGER1a. However, biochemical experiments are consistent with the predicted network phenotype. The ability of networks to predict counter-intuitive *in vivo* phenotypes for mutant/knock-out cells may not always hold true, as we have observed it only for one node in our network. Nevertheless, this idea has appeal. It is not unprecedented that biochemical interactions under cellular conditions, and in presence of other interacting biochemical components can behave differently from *in vitro* interactions. Given this, modeling studies should also use a holistic approach, and study cell as a system and not just as reductionist version of individual interactions. Our network illustrates that this systems biological perspective for studying cells as a network of cross-interacting biological reactions, rather than as individual interactions, may provide a more accurate depiction of cellular processes.

To date, Boolean analysis has been utilized to study gene regulation patterns[47-49], Abscisic acid signaling pathway [29], immune responses[50], cholesterol biosynthesis [51], and survival signaling in T-cells[52]. Taken together with our analysis, we theorize that Boolean network can be used to study any biological pathways, providing clues for further experimental work. After experimental verification, this new information can feed back into the network, thereby improving the model.

A limiting factor to the Boolean modeling approach is that it depends on information obtained from different sources. Thus, the quality and accuracy of information used to define logical rules can significantly impact the network results. Likewise bionetworks can be incomplete, as in the case of this study. There are many other calcium networks described (but not modeled) which are bigger and more complex (e.g. Berridge model [53]), but in this paper we have focused only on a smaller network that forms backbone of all GPCR and RTK calcium networks. However, novel nodes and edges will be added in future, as our understanding of these pathways gets better. It is hard to say how much of the results obtained from this study might change after addition of novel nodes and edges in the future. Nonetheless, these experimental data can be used to iteratively define the rules for our Boolean modeling, with the resulting predictions paving the way for further experimental research.

The network presented here can be accessed online, and codes for creating Boolean network models are available at BooleanNet (http://code.google.com/p/booleannet/). We encourage the $Ca^{2+}$-signaling community to provide additional rules, which have been experimentally validated, to be incorporated into these networks. Currently, our bionetwork is very specific; however, the modular nature of Boolean networks makes it very easy to modify and/or increase the network complexity. Thus, we envision that development of this resource will likely lead to the construction of cell-specific $Ca^{2+}$ signaling networks. Further, we believe that this wiki approach will allow for an accurate and detailed expansion of the $Ca^{2+}$ signaling network, which in turn will improve its predictive power.

In congress with two companion studies (Hong et al, Ko et al, Physics Archives 2009), we theorize that phylogenetic profiles can be used to improve network connectivity. Indeed, phylogenetic profiles are robust in predicting protein functionality, even in highly divergent sequences. As these predictions of functionality can be encoded into Boolean operators, it stands to reason that this information would improve the predictive performance of the network. Further, other predictions that can be obtained from phylogenetic profiles, such as protein-protein interactions, can also aid in creating additional edges between nodes. It is reasonable to consider that networks incorporating this data may also be useful in identifying rogue predictions from our phylogenetic profiles. For instance, if protein A is predicted to be a kinase that phosphorylates protein B, but by incorporating this rule the network no longer adheres to the known relationships, it calls into question the validity of the phylogenetic profile prediction. This is an attractive iterative process as Boolean networks are computationally inexpensive. To what extent this and other types of predictive computational biology can be incorporated into Boolean networks, and their efficacy, is an exciting realm for exploration.

In the future, our work will also focus on developing a more complex model by including nodes and connections already described in Berridge model [54] and by including other phenotypic characteristics into the bionetwork. For example, DANGER1a, TRPCs, and $IP_3R$ are also involved in neural growth [4;44;55-57]. Therefore, a neuronal development network can be integrated into the $Ca^{2+}$ signaling network. In theory, this should allow for the simultaneous decoding of these two highly interconnected networks. Since calcium mobilization is not an ON/OFF process, we will also explore higher order discrete modeling approaches towards better simulations. We will also test hybrids of discrete/continuous models to better understand calcium mobilization; particularly in cases where some kinetic data is available. In addition to modeling signaling pathways, these networks can aid in the identification of drug targets based on network connectivity (i.e. some nodes serve as better drug targets than others [58;59]). We propose that in the long-term, our refined Boolean modeling may lead to increased capacity to create predictive stage models likely to identify targets for therapeutic intervention.

**Acknowledgments**
This work was supported by the Searle Young Investigators Award and start-up money from PSU (RLP), NCSA grant TG-MCB070027N (RLP, DVR), the T.O. Magnetti Fund (RLP,DVR), The National Science Foundation 428-15 691M (RLP, DVR), and The National Institutes of Health R01 GM087410-01 (RLP, DVR). This project was also funded by a Fellowship from the Eberly College of Sciences and the Huck Institutes of the Life Sciences (DVR) and a grant with the Pennsylvania Department of Health using Tobacco Settlement Funds (DVR). The Department of Health specifically disclaims responsibility for any analyses, interpretations or conclusions. We would like to thank Sunando Roy, Andre Wallace, Gue Su Chang, Yoojin Hong, Kyung Dae Ko for their help and support during the project. We would also like to thank Jason Holmes at the Pennsylvania State University CAC center for technical assistance and Drs. Boo Rothe, J. Goldhammer, F. Flave, Sweets Vollmar, T. Nerds, Y.G. Tea, T. Comish, and D. Octagon for creative dialogue.

**Materials and Methods**
**Bionetwork and Boolean Modeling**
Nodes and edges in bionetwork were chosen after a comprehensive literature search for components that are part of Serotonin $Ca^{2+}$ pathway and/or are known to interact with any of the other components of the network. Some of the interactions are very well established based on experimental data but in cases where conflicting results existed, nodes and edges were defined based on relevance and accuracy of experimental support provided in literature. Further, only nodes that have causal effects on the system were included. Associative interactions, where directionality information was not present in current literature could not be included. For a list of references used in generating the network, see Supplemental References.

The model in Figure 1 was drawn with yED graph editor (http://www.yworks.com/en/index.html). The Boolean modeling was performed with a custom Python script (http://www.python.org). Update rules for each node were defined based on legacy knowledge and experimental data obtained prior to this modeling.(Table1). For synchronous modeling, all the nodes were updated for their respective rules at all the time steps. For asynchronous modeling, random numbers of nodes were updated at each time step. A random update order was selected at each step from N! possible permutations (N = no. of nodes). As shown in Chaves et al and Li et al [29;60], this method provides a random differential time to each reaction/interaction event. Boolean states of each node at each step (for 50 time steps in case of synchronous) were recorded and analyzed for repeating patterns (Attracting behavior).

For random rewiring of AND and OR in logical rules, all the ANDs in the logical rules were changed to OR and vice versa. Synchronous update method was used after rewiring to get an equivalent of randomization in this network study.

**Calcium Imaging-** Calcium imaging was performed as previously described [61]. Briefly neurons were loaded with 1mM Fura-2-AM for 5 minutes. B-cells were loaded with 2mM Fura-2-AM for 25 minutes. Neuronal experiments were conducted at $37^0$C and Kreb's solution was supplemented with 2mM $CaCl_2$ and 1mM glycine. B-cell experiments were conducted at room temperature as previously described [32]. HEK-293 and A7r5 cells were imaged exactly as in Patterson et al [32].

**Cell Culture-** Cells were cultured using the protocol from Alldred et al [62]. Briefly, spinal cord neurons were generated from DANGER1A deficient embryonic day 14.5 embryos generated by crossing of DANGER1A$^{+/-}$ mice on a 129SvJ inbred background. Spinal cords were collected in PBS containing 5.5 mM glucose and treated with papain (0.5 mg/ml) and DNase I (10 µg/ml) (both from Sigma, St. Louis, MO) in PBS containing 1 mg/ml bovine serum albumin (Fraction V, Sigma) and 10 mM glucose for 15 min at room temperature. The cells were triturated with a fire-polished Pasteur pipette and plated on poly-L-lysine-coated glass coverslips (22 x 22 mm) at 4 x $10^4$ cells per square centimeter in modified Eagle medium (MEM) (Invitrogen) containing 10% v/v fetal bovine serum (FBS) (Invitrogen) in an atmosphere of 10% $CO_2$. After 60 min, the medium was replaced with fresh MEM containing 10% v/v FBS. The genotype of cultures was determined using PCR as follows. Tail biopsies (3 mm) were incubated for 30 min at 55°C in 25 µl lysis buffer (200 mM NaCl, 5 mM EDTA, 0.2% SDS, 100 mM Tris-HCl, pH 8.5), and 1% of the supernatant was used for PCR using standard conditions with the primer 5'-ACAGAACCAATCGGTGTG-3' combined with either 5'-CTCGTTCTCGGGAATCGT-3' to amplify the wild-type DANGER1A locus or the primer 5'-GCGCTTCCAAGGCTGTGAA-3' to amplify the mutant DANGER1A locus. The 24-hr-old cultures that exhibited the desired genotypes were turned upside down onto a glial feeder layer in a Petri dish containing Neurobasal-A supplemented with B27 (Invitrogen), in an atmosphere of 10% $CO_2$. Feeder cells were prepared from cortices of newborn rat pups as described (Banker and Goslin, 1998+). Neuron cultures were maintained without medium change for 18 d *in vitro* (DIV) and then transferred into new Petri dishes containing Neurobasal A/B27 supplemented with 1 µM 6-cyano-7-nitroquinoxaline-2,3-dione (CNQX) and 100 µM 2-amino-5-phosphonovaleric acid (Sigma), with the cells facing up. Neurons were processed for calcium imaging at 10 DIV.

**Microsomal Assays-** Microsomal assays were performed as previously described [63;64].
***In vitro* Binding assays –** *In vitro* assays were performed exactly as previously described [63;65]

**Yeast-2-Hybrid Assays** – Yeast-2-Hybrid assays were performed exactly as previously described [63;65].

**Materials-** Anti-myc antibody, 5-HT, Anti-human IgM, carbachol, and all buffer components were obtained from Sigma-Aldrich (St. Louis, MO). Yeast-2-Hybrid reagents were obtained from CLONETECH (Palo Alto, CA). All cell culture reagents and Fura-2-AM were obtained from Invitrogen. $^3$IP$_3$ and $^{45}$Ca$^{2+}$ were obtained from Amersham Biosciences. DANGER1a knock-out mice were generated by Ozgene© (Australia).

# References

[1] R. L. Patterson, D. Boehning, and S. H. Snyder, "Inositol 1,4,5-trisphosphate receptors as signal integrators," Annu. Rev Biochem., vol. 73, pp. 437-465, 2004.

[2] P. Pinton, C. Giorgi, R. Siviero, E. Zecchini, and R. Rizzuto, "Calcium and apoptosis: ER-mitochondria Ca2+ transfer in the control of apoptosis," Oncogene, vol. 27, no. 50, pp. 6407-6418, Oct.2008.

[3] M. J. Berridge, P. Lipp, and M. D. Bootman, "The versatility and universality of calcium signalling," Nat. Rev. Mol. Cell Biol., vol. 1, no. 1, pp. 11-21, Oct.2000.

[4] K. Venkatachalam and C. Montell, "TRP channels," Annu. Rev. Biochem., vol. 76, pp. 387-417, 2007.

[5] K. Venkatachalam, F. Zheng, and D. L. Gill, "Regulation of canonical transient receptor potential (TRPC) channel function by diacylglycerol and protein kinase C," J. Biol. Chem., vol. 278, no. 31, pp. 29031-29040, Aug.2003.

[6] D. E. Clapham, "TRP channels as cellular sensors," Nature, vol. 426, no. 6966, pp. 517-524, Dec.2003.

[7] H. S. Li, X. Z. Xu, and C. Montell, "Activation of a TRPC3-dependent cation current through the neurotrophin BDNF," Neuron, vol. 24, no. 1, pp. 261-273, Sept.1999.

[8] S. Shim, E. L. Goh, S. Ge, K. Sailor, J. P. Yuan, H. L. Roderick, M. D. Bootman, P. F. Worley, H. Song, and G. L. Ming, "XTRPC1-dependent chemotropic guidance of neuronal growth cones," Nat Neurosci., vol. 8, no. 6, pp. 730-735, June2005.

[9] D. P. Corey, J. Garcia-Anoveros, J. R. Holt, K. Y. Kwan, S. Y. Lin, M. A. Vollrath, A. Amalfitano, E. L. Cheung, B. H. Derfler, A. Duggan, G. S. Geleoc, P. A. Gray, M. P. Hoffman, H. L. Rehm, D. Tamasauskas, and D. S. Zhang, "TRPA1 is a candidate for the mechanosensitive transduction channel of vertebrate hair cells," Nature, vol. 432, no. 7018, pp. 723-730, Dec.2004.

[10] K. Tabuchi, M. Suzuki, A. Mizuno, and A. Hara, "Hearing impairment in TRPV4 knockout mice," Neurosci. Lett., vol. 382, no. 3, pp. 304-308, July2005.

[11] Y. Zhang, M. A. Hoon, J. Chandrashekar, K. L. Mueller, B. Cook, D. Wu, C. S. Zuker, and N. J. Ryba, "Coding of sweet, bitter, and umami tastes: different receptor cells sharing similar signaling pathways," Cell, vol. 112, no. 3, pp. 293-301, Feb.2003.

[12] M. K. Jungnickel, H. Marrero, L. Birnbaumer, J. R. Lemos, and H. M. Florman, "Trp2 regulates entry of Ca2+ into mouse sperm triggered by egg ZP3," Nat. Cell Biol., vol. 3, no. 5, pp. 499-502, May2001.

[13] K. Kiselyov, A. Soyombo, and S. Muallem, "TRPpathies," J. Physiol, vol. 578, no. Pt 3, pp. 641-653, Feb.2007.

[14] M. T. Miedel, Y. Rbaibi, C. J. Guerriero, G. Colletti, K. M. Weixel, O. A. Weisz, and K. Kiselyov, "Membrane traffic and turnover in TRP-ML1-deficient cells: a revised model for mucolipidosis type IV pathogenesis," J. Exp. Med., May2008.

[15] S. E. Jordt and B. E. Ehrlich, "TRP channels in disease," Subcell. Biochem., vol. 45, pp. 253-271, 2007.

[16] Y. X. Li, "Equations for InsP3 receptor-mediated[Ca2+]i oscillations derived from a detailed kinetic model: a Hodgkin-Huxley like formalism," Journal of theoretical biology, vol. 166, no. 4, p. 461, 1994.

[17] M. R. Maurya and S. Subramaniam, "A kinetic model for calcium dynamics in RAW 264.7 cells: 1. Mechanisms, parameters, and subpopulational variability," Biophys. J., vol. 93, no. 3, pp. 709-728, Aug.2007.

[18] M. R. Maurya and S. Subramaniam, "A kinetic model for calcium dynamics in RAW 264.7 cells: 2. Knockdown response and long-term response," Biophys. J., vol. 93, no. 3, pp. 729-740, Aug.2007.

[19] R. H. Chow, "Cadmium block of squid calcium currents. Macroscopic data and a kinetic model," The Journal of General Physiology, vol. 98, no. 4, pp. 751-770, Oct.1991.

[20] Y. Tang, "Simplification and analysis of models of calcium dynamics based on IP3-sensitive calcium channel kinetics," Biophys. J., vol. 70, no. 1, p. 246, 1996.

[21] U. S. Bhalla and R. Iyengar, "Emergent properties of networks of biological signaling pathways," Science, vol. 283, no. 5400, pp. 381-387, Jan.1999.

[22] G. von Dassow, "The segment polarity network is a robust developmental module," Nature, vol. 406, no. 6792, p. 188, 2000.

[23] D. H. Sharp and J. Reinitz, "Prediction of mutant expression patterns using gene circuits," Biosystems, vol. 47, no. 1-2, pp. 79-90, June1998.

[24] M. Chaves, E. D. Sontag, and R. Albert, "Methods of robustness analysis for Boolean models of gene control networks," Syst. Biol (Stevenage. ), vol. 153, no. 4, pp. 154-167, July2006.

[25] R. Albert and H. G. Othmer, "The topology of the regulatory interactions predicts the expression pattern of the segment polarity genes in Drosophila melanogaster," Journal of theoretical biology, vol. 223, no. 1, pp. 1-18, July2003.

[26] R. Albert, "Scale-free networks in cell biology," J. Cell Sci, vol. 118, no. Pt 21, pp. 4947-4957, Nov.2005.

[27] M. Chaves, R. Albert, and E. D. Sontag, "Robustness and fragility of Boolean models for genetic regulatory networks," J. Theor. Biol., vol. 235, no. 3, pp. 431-449, Aug.2005.

[28] R. L. Patterson, D. B. van Rossum, N. Nikolaidis, D. L. Gill, and S. H. Snyder, "Phospholipase C-gamma: diverse roles in receptor-mediated calcium signaling," Trends Biochem. Sci, vol. 30, no. 12, pp. 688-697, Dec.2005.

[29] S. Li, S. M. Assmann, and R. Albert, "Predicting Essential Components of Signal Transduction Networks: A Dynamic Model of Guard Cell Abscisic Acid Signaling," PLoS. Biol, vol. 4, no. 10 Sept.2006.

[30] W. Zeng, D. O. Mak, Q. Li, D. M. Shin, J. K. Foskett, and S. Muallem, "A new mode of Ca2+ signaling by G protein-coupled receptors: gating of IP3 receptor Ca2+ release channels by Gbetagamma," Curr. Biol, vol. 13, no. 10, pp. 872-876, May2003.

[31] V. Ruiz-Velasco and S. R. Ikeda, "A splice variant of the G protein beta 3-subunit implicated in disease states does not modulate ion channels," Physiol Genomics, vol. 13, no. 2, pp. 85-95, Apr.2003.

[32] R. L. Patterson, D. B. van Rossum, D. L. Ford, K. J. Hurt, S. S. Bae, P. G. Suh, T. Kurosaki, S. H. Snyder, and D. L. Gill, "Phospholipase C-gamma is required for agonist-induced Ca2+ entry," Cell, vol. 111, no. 4, pp. 529-541, Nov.2002.

[33]   M. R. Clark, K. S. Campbell, A. Kazlauskas, S. A. Johnson, M. Hertz, T. A. Potter, C. Pleiman, and J. C. Cambier, "The B cell antigen receptor complex: association of Ig-alpha and Ig-beta with distinct cytoplasmic effectors," Science, vol. 258, no. 5079, pp. 123-126, Oct.1992.

[34]   A. Tordai, R. A. Franklin, H. Patel, A. M. Gardner, G. L. Johnson, and E. W. Gelfand, "Cross-linking of surface IgM stimulates the Ras/Raf-1/MEK/MAPK cascade in human B lymphocytes," J. Biol. Chem., vol. 269, no. 10, pp. 7538-7543, Mar.1994.

[35]   C. A. Ross, S. K. Danoff, M. J. Schell, S. H. Snyder, and A. Ullrich, "Three additional inositol 1,4,5-trisphosphate receptors: molecular cloning and differential localization in brain and peripheral tissues," Proc. Natl. Acad. Sci. U. S. A, vol. 89, no. 10, pp. 4265-4269, May1992.

[36]   A. H. Sharp, S. H. Snyder, and S. K. Nigam, "Inositol 1,4,5-trisphosphate receptors. Localization in epithelial tissue," J. Biol. Chem., vol. 267, no. 11, pp. 7444-7449, Apr.1992.

[37]   A. P. Dawson, "Calcium signalling: How do IP3 receptors work?," Curr. Biol., vol. 7, no. 9, p. R544-R547, 1997.

[38]   V. Ruiz-Velasco and S. R. Ikeda, "Multiple G-Protein beta gamma Combinations Produce Voltage-Dependent Inhibition of N-Type Calcium Channels in Rat Superior Cervical Ganglion Neurons," J. Neurosci., vol. 20, no. 6, pp. 2183-2191, Mar.2000.

[39]   M. J. Berridge, M. D. Bootman, and H. L. Roderick, "Calcium signalling: dynamics, homeostasis and remodelling," Nat Rev Mol Cell Biol, vol. 4, no. 7, pp. 517-529, July2003.

[40]   J. Y. Kim, W. Zeng, K. Kiselyov, J. P. Yuan, M. H. Dehoff, K. Mikoshiba, P. F. Worley, and S. Muallem, "Homer 1 mediates store- and inositol 1,4,5-trisphosphate receptor-dependent translocation and retrieval of TRPC3 to the plasma membrane," J. Biol. Chem., vol. 281, no. 43, pp. 32540-32549, Oct.2006.

[41]   J. P. Yuan, K. I. Kiselyov, D. M. Shin, J. Chen, N. Shcheynikov, H. S. Kang, M. H. Dehoff, M. K. Schwarz, P. H. Seeburg, S. Muallem, and P. F. Worley, "Homer Binds TRPC Family Channels and Is Required for Gating of TRPC1 by IP3-Receptors," Cell, vol. 114, pp. 777-789, 2003.

[42]   H. Ando, A. Mizutani, T. Matsu-ura, and K. Mikoshiba, "IRBIT, a novel inositol 1,4,5-trisphosphate (IP3) receptor-binding protein, is released from the IP3 receptor upon IP3 binding to the receptor," J. Biol. Chem., vol. 278, no. 12, pp. 10602-10612, Mar.2003.

[43]   H. Ando, A. Mizutani, H. Kiefer, D. Tsuzurugi, T. Michikawa, and K. Mikoshiba, "IRBIT suppresses IP3 receptor activity by competing with IP3 for the common binding site on the IP3 receptor," Mol Cell, vol. 22, no. 6, pp. 795-806, June2006.

[44]   N. Nikolaidis, D. Chalkia, D. N. Watkins, R. K. Barrow, S. H. Snyder, D. B. van Rossum, and R. L. Patterson, "Ancient Origin of the New Developmental Superfamily DANGER," PLoS. ONE., vol. 2, p. e204, 2007.

[45]   D. B. van Rossum, R. L. Patterson, K. H. Cheung, R. K. Barrow, V. Syrovatkina, G. S. Gessell, S. G. Burkholder, D. N. Watkins, J. K. Foskett, and S. H. Snyder, "DANGER: A novel regulatory protein of IP3-receptor activity," J. Biol Chem., Sept.2006.

[46]   J. Soboloff, M. Spassova, T. Hewavitharana, L. P. He, P. Luncsford, W. Xu, K. Venkatachalam, R. D. van, R. L. Patterson, and D. L. Gill, "TRPC channels: integrators of multiple cellular signals," Handb. Exp. Pharmacol., no. 179, pp. 575-591, 2007.

[47]   C. Espinosa-Soto, P. Padilla-Longoria, and E. R. varez-Buylla, "A Gene Regulatory Network Model for Cell-Fate Determination during Arabidopsis thaliana Flower Development That Is Robust and Recovers Experimental Gene Expression Profiles," Plant Cell, vol. 16, no. 11, pp. 2923-2939, Nov.2004.

[48] D. Liu, D. M. Umbach, S. D. Peddada, L. Li, P. W. Crockett, and C. R. Weinberg, "A random-periods model for expression of cell-cycle genes," Proc. Natl. Acad. Sci. USA, vol. 101, no. 19, pp. 7240-7245, May2004.

[49] J. Thakar, M. Pilione, G. Kirimanjeswara, E. T. Harvill, and R. Albert, "Modeling systems-level regulation of host immune responses," PLoS. Comput. Biol., vol. 3, no. 6, p. e109, June2007.

[50] G. Kervizic and L. Corcos, "Dynamical modeling of the cholesterol regulatory pathway with Boolean networks," BMC. Syst. Biol., vol. 2, p. 99, 2008.

[51] R. Zhang, M. V. Shah, J. Yang, S. B. Nyland, X. Liu, J. K. Yun, R. Albert, and T. P. Loughran, "Network model of survival signaling in large granular lymphocyte leukemia," Proceedings of the National Academy of Sciences, vol. 105, no. 42, pp. 16308-16313, Oct.2008.

[52] K. Takei, R. M. Shin, T. Inoue, K. Kato, and K. Mikoshiba, "Regulation of nerve growth mediated by inositol 1,4,5-trisphosphate receptors in growth cones," Science, vol. 282, no. 5394, pp. 1705-1708, Nov.1998.

[53] K. Venkatachalam, H. T. Ma, D. L. Ford, and D. L. Gill, "Expression of functional receptor-coupled TRPC3 channels in DT40 triple receptor InsP3 knockout cells," J. Biol. Chem., vol. 276, no. 36, pp. 33980-33985, Sept.2001.

[54] Y. Li, Y. C. Jia, K. Cui, N. Li, Z. Y. Zheng, Y. Z. Wang, and X. B. Yuan, "Essential role of TRPC channels in the guidance of nerve growth cones by brain-derived neurotrophic factor," Nature, vol. 434, no. 7035, pp. 894-898, Apr.2005.

[55] H. Jeong, Z. N. Oltvai, and A. L. Barabbsi, "Prediction of Protein Essentiality Based on Genomic Data," Complexus, vol. 1, no. 1, pp. 19-28, 2003.

[56] B. M. Rao, D. A. Lauffenburger, and K. D. Wittrup, "Integrating cell-level kinetic modeling into the design of engineered protein therapeutics," Nat Biotech, vol. 23, no. 2, pp. 191-194, Feb.2005.

[57] M. Chaves, R. Albert, and E. D. Sontag, "Robustness and fragility of Boolean models for genetic regulatory networks," J. Theor. Biol., vol. 235, no. 3, pp. 431-449, Aug.2005.

[58] D. B. van Rossum, D. Oberdick, Y. Rbaibi, G. Bhardwaj, R. K. Barrow, N. Nikolaidis, S. H. Snyder, K. Kiselyov, and R. L. Patterson, "TRP_2, a Lipid/Trafficking Domain That Mediates Diacylglycerol-induced Vesicle Fusion," J. Biol. Chem., vol. 283, no. 49, pp. 34384-34392, Dec.2008.

[59] M. J. Alldred, J. Mulder-Rosi, S. E. Lingenfelter, G. Chen, and B. Luscher, "Distinct gamma2 subunit domains mediate clustering and synaptic function of postsynaptic GABAA receptors and gephyrin," J. Neurosci., vol. 25, no. 3, pp. 594-603, Jan.2005.

[60] R. L. Patterson, D. B. van Rossum, R. K. Barrow, and S. H. Snyder, "RACK1 binds to inositol 1,4,5-trisphosphate receptors and mediates Ca2+ release," Proc. Natl. Acad. Sci. U. S. A, vol. 101, no. 8, pp. 2328-2332, Feb.2004.

[61] D. Boehning, R. L. Patterson, L. Sedaghat, N. O. Glebova, T. Kurosaki, and S. H. Snyder, "Cytochrome c binds to inositol (1,4,5) trisphosphate receptors, amplifying calcium-dependent apoptosis," Nat. Cell Biol., Nov.2003.

[62] R. L. Patterson, D. B. van Rossum, A. I. Kaplin, R. K. Barrow, and S. H. Snyder, "Inositol 1,4,5-trisphosphate receptor/GAPDH complex augments Ca2+ release via locally derived NADH," Proc. Natl. Acad Sci U. S. A, vol. 102, no. 5, pp. 1357-1359, Feb.2005.

[63] H. Streb, R. F. Irvine, M. J. Berridge, and I. Schulz, "Release of Ca2+ from a nonmitochondrial intracellular store in pancreatic acinar cells by inositol-1,4,5-trisphosphate," Nature, vol. 306, no. 5938, pp. 67-69, Nov.1983.

[64] M. Trebak, N. Hempel, B. J. Wedel, J. T. Smyth, G. S. Bird, and J. W. Putney, Jr., "Negative regulation of TRPC3 channels by protein kinase C-mediated phosphorylation of serine 712," Mol. Pharmacol., vol. 67, no. 2, pp. 558-563, Feb.2005.

[65] C. D. Ferris, R. L. Huganir, and S. H. Snyder, "Calcium flux mediated by purified inositol 1,4,5-trisphosphate receptor in reconstituted lipid vesicles is allosterically regulated by adenine nucleotides," Proc. Natl. Acad. Sci. U. S. A, vol. 87, no. 6, pp. 2147-2151, Mar.1990.

[66] D. O. Mak, S. McBride, and J. K. Foskett, "ATP regulation of type 1 inositol 1,4,5-trisphosphate receptor channel gating by allosteric tuning of Ca(2+) activation," J. Biol. Chem., vol. 274, no. 32, pp. 22231-22237, Aug.1999.

[67] L. P. Haynes, A. V. Tepikin, and R. D. Burgoyne, "Calcium binding protein 1 is an inhibitor of agonist-evoked, inositol 1,4,5-trisphophate-mediated calcium signalling," J. Biol. Chem., p. M309617200, Oct.2003.

[68] T. Uchiyama, F. Yoshikawa, A. Hishida, T. Furuichi, and K. Mikoshiba, "A Novel Recombinant Hyperaffinity Inositol 1,4,5-Trisphosphate (IP3) Absorbent Traps IP3, Resulting in Specific Inhibition of IP3-mediated Calcium Signaling," J. Biol. Chem., vol. 277, no. 10, pp. 8106-8113, Mar.2002.

[69] J. I. Bruce, T. J. Shuttleworth, D. R. Giovannucci, and D. I. Yule, "Phosphorylation of inositol 1,4,5-trisphosphate receptors in parotid acinar cells. A mechanism for the synergistic effects of cAMP on Ca2+ signaling," J. Biol. Chem., vol. 277, no. 2, pp. 1340-1348, Jan.2002.

[70] S. V. Straub, D. R. Giovannucci, J. I. Bruce, and D. I. Yule, "A role for phosphorylation of inositol 1,4,5-trisphosphate receptors in defining calcium signals induced by Peptide agonists in pancreatic acinar cells," J. Biol. Chem., vol. 277, no. 35, pp. 31949-31956, Aug.2002.

[71] M. Grimaldi, A. Favit, and D. L. Alkon, "cAMP-induced cytoskeleton rearrangement increases calcium transients through the enhancement of capacitative calcium entry," J. Biol. Chem., vol. 274, no. 47, pp. 33557-33564, Nov.1999.

[72] P. J. Bartlett, K. W. Young, S. R. Nahorski, and R. A. Challiss, "Single cell analysis and temporal profiling of agonist-mediated inositol 1,4,5-trisphosphate, Ca2+, diacylglycerol, and protein kinase C signaling using fluorescent biosensors," J. Biol. Chem., vol. 280, no. 23, pp. 21837-21846, June2005.

[73] U. H. Kim, D. Fink, Jr., H. S. Kim, D. J. Park, M. L. Contreras, G. Guroff, and S. G. Rhee, "Nerve growth factor stimulates phosphorylation of phospholipase C-gamma in PC12 cells," J. Biol. Chem., vol. 266, no. 3, pp. 1359-1362, Jan.1991.

[74] J. K. Acharya, K. Jalink, R. W. Hardy, V. Hartenstein, and C. S. Zuker, "InsP3 receptor is essential for growth and differentiation but not for vision in Drosophila," Neuron, vol. 18, no. 6, pp. 881-887, June1997.

[75] M. J. Rebecchi and S. N. Pentyala, "Structure, function, and control of phosphoinositide-specific phospholipase C," Physiol Rev., vol. 80, no. 4, pp. 1291-1335, Oct.2000.

[76] S. G. Rhee, "Regulation of phosphoinositide-specific phospholipase C," Annu. Rev. Biochem., vol. 70, pp. 281-312, 2001.

[77] C. L. Tu, W. Chang, and D. D. Bikle, "Phospholipase cgamma1 is required for activation of store-operated channels in human keratinocytes," J. Invest Dermatol., vol. 124, no. 1, pp. 187-197, Jan.2005.

[78] S. Patel, S. K. Joseph, and A. P. Thomas, "Molecular properties of inositol 1,4,5-trisphosphate receptors," Cell Calcium, vol. 25, no. 3, pp. 247-264, Mar.1999.

[79] T. Michikawa, J. Hirota, S. Kawano, M. Hiraoka, M. Yamada, T. Furuichi, and K. Mikoshiba, "Calmodulin mediates calcium-dependent inactivation of the cerebellar type 1 inositol 1,4,5-trisphosphate receptor," Neuron, vol. 23, no. 4, pp. 799-808, Aug.1999.

[80] L. Missiaen, J. B. Parys, A. F. Weidema, H. Sipma, S. Vanlingen, P. De Smet, G. Callewaert, and H. De Smedt, "The bell-shaped Ca2+ dependence of the inositol 1,4, 5-trisphosphate-induced Ca2+ release is modulated by Ca2+/calmodulin," J. Biol. Chem., vol. 274, no. 20, pp. 13748-13751, May1999.